\documentclass{article}
\usepackage{amssymb}

\usepackage{graphicx}
\usepackage{amsmath}

\begin{document}

\title{\textbf{Fantappi\`{e}'s group as an extension of special relativity on $\epsilon ^{(\infty )}$ Cantorian
space-time}}
\author{G.Iovane\thanks{%
iovane@diima.unisa.it}, P.Giordano \\
%EndAName
Dipartimento di Ingegneria dell'Informazione e Matematica Applicata,\\
Universit\'a di Salerno, Italy.\\
\\
E.Laserra\\
%EndAName
Dipartimento di Matematica e Informatica\\
Universit\'a di Salerno, Italy.}
\date{31.03.2004}
\maketitle

\begin{abstract}
In this paper we will analyze the Fantappi\`{e} group and its
properties in connection with Cantorian space-time. Our attention
will be focused on the possibility of extending special
relativity. The cosmological consequences of such extension appear
relevant, since thanks to the Fantappi\`{e} group, the model of
the Big Bang and that of stationary state become compatible. In
particular, if we abandon the idea of the existence of only one
time gauge, since we do not see the whole Universe but only a
projection, the two models become compatible. In the end we will
see the effects of the projective fractal geometry also on the
galactic and extra--galactic dynamics.
\end{abstract}

%\newpage

\section{Introduction}
It is known that invariance under Lorentz transformations is a
fundamental principle underlying both relativity and quantum field
theory. Recently it has been suggested that global Lorentz
invariance is an approximation of nature, that can be broken at
high-energy physics. Also in the detection of ultrahigh energy
cosmic rays and TeV photon spectra some anomalies are found
\cite{Bech}, \cite{Amelino1}, \cite{Amelino2}. Many authors showed
that vacuum fluctuations and quantum gravity effects introduce
stochastic perturbations in the space-time geometry at Plank
energy scale \cite{Aloisio}, \cite{Bertolami}, \cite{Bech},
\cite{Amelino1}, \cite{Coleman}, \cite{Elnaschie10},
\cite{Elnaschie11}, \cite{Elnaschie12}, \cite{Goldfain}. In
\cite{Iovane4} G.Iovane showed the relevant consequences of a
Stochastic Self-Similar and Fractal Universe. Starting from an
universal scaling law, the author showed its agreement with the
well--known Random Walk equation or
Brownian motion relation that was used by Eddington \cite{Elnaschie3}, \cite%
{Sidharth}. Consequently, he arrived at a self-similar Universe.
It appears that the Universe has a memory of its quantum origin as
suggested by R.Penrose with respect to quasi-crystal
\cite{Penrose}.
Particularly, the model was related to Penrose tiling and thus to $%
\varepsilon ^{(\infty )}$ theory (Cantorian space-time theory) as
proposed by El Naschie \cite{Elnaschie1},\cite{Elnaschie2} as well
as with Connes Noncommutative Geometry \cite{Connes}. In
\cite{Iovane5} the authors presented a descriptive model of
segregated universe, then considered a dynamical model to explain
the results and to give the evolution of the structures.

It is already well known that cosmology can be analyzed thanks to
different theories which are not always compatible with each
other. Actually the major part of the theories of gravitation are
obtained by modifying the Einstein-Hilbert action, adding scalar
fields or curvature invariants in the form $\phi
^{2}R,R^{2},R_{\mu \nu }R^{\mu \nu }$ or $R\square R$ (\cite
{Weinberg}, \cite{Starobinsky}, \cite{Eckhardt}). If we apply
Einstein's equations to the whole Universe, we find the
relativistic cosmology, in which the cosmological principle can be
postulated and a model of constant spatial curvature obtained. We
have to pay close attention to general relativity, where,
inevitably, the application of Einstein's equations to
cosmological problems requires an extreme extrapolation of their
validity to very far regions of space-time. Therefore we can look
for solutions which present a cosmological interest, as long as we
take into account, for this type of problem, such equations could
be little more than a good model. As it is known, minor changes to
the equations, while exhibiting all the classical verifications,
produce completely different cosmologically--interesting solutions
\cite{Ruffini}, \cite{Kolb}. It is useful to observe that even
though cosmology accepts general relativity as a definitive theory
of gravitation, there are still some uncertain aspects due to a
baffling pluralism. Possible universes are numerous and differ
from each other substantially. Moreover some astronomers, such as
Arp and Hoyle, believe that to connect the red shift to the
recession is an error because it is known there are other
mechanisms which produce the red shift. Arp affirms that some
astronomical objects appear to be gravitationally interacting
among themselves, and so they should be spatially near. Instead
their red shift indicates very different velocities of recession.
In addition there are some objects which appear to be older than
the Universe and, for this reason, Arp has proposed to return to a
variation of the old stationary model in which there isn't an
origin of time \cite{Arp0}, \cite{Arp1}, \cite{Arp2}. According to
this theory, formulated by Bondi, Gold and Hoyle, the Universe has
always been as we see it today. Concerning the theory of the
Big-Bang, the primary difficulty is the presence of the initial
singularity which brings up the problem of behavior of matter when
it is reduced to no dimensions with infinite density and
temperature. Many difficulties of the Big Bang standard model can
be overcome by inflationary cosmology \cite{Linde}, \cite{Israel},
\cite{Narlikar}. It is actually difficult to find a theory that
explains what Big-Bang really looked like. Probably, many think
that all this will be resolved when we are able to formulate a
quantum theory of gravitation. As it is known, Hawking and Hartle
have looked for a way, based on quantum mechanics, to explain how
time could have spontaneously begun in correspondence with
Big--Bang \cite{Hawking1}. The idea is that time could have been
imaginary, similar to space, near Big--Bang \cite{Hawking2}. That
is, in proximity to Big-Bang, it would be more exact to speak of
4--dimensional space instead of space--time. The hypothesis that
the Universe had its origin in a singularity of infinite
compression can be graphically represented by a cone with its
point at the base of the diagram. As it is well known, in the
quantum cosmology of Hartle and Hawking the point of the cone is
substituted by a half--sphere with a radius equal to the length of
Planck, $10^{-33}cm$. In the upper part of this half--sphere the
cone widens in the usual way representing the standard development
of the Universe in expansion. The transition from imaginary time
to real time is gradual and cannot be proven all at once. The
conclusion is that according to this approach, there is no origin
of the Universe even if time is limited in the past. In
\cite{Elnaschie13} Mohamed El Naschie also consider the imaginary
time and seek a formal definition of nowness. Starting from this
result De Felice et al, consider Lorentz transformations and
complex space-time functions \cite{Defelice}.

In this paper we want to show that, even if Einstein's ideas
triggered a revolutionary process in our comprehension of space
and time, the relativistic space--time could not be sufficient to
completely explain the physical Universe and our perception of it.
Expanding the idea of Hartle and Hawking to the whole space--time
manifold, we find ourselves in a model of the Universe in which
geometry is linked to the group that Fantappi\`{e} obtained by
generalizing Poincar\`{e}'s relativistic group. Thus, according to
the official interpretation of quantum mechanics, the observer
plays a fundamental role in the description of the atomic world.
Analogously, also if in a different way, we think that in the
description of the universe, the observer becomes more deeply
involved in respect to what is generally believed. For this reason
we have made a distinction between space--time, external to the
observer and associated with gravitation, and the internal
space--time, associated with the other interactions. This last one
is regulated by Fantappi\`{e}'s transformations. A relevant point
is that this space-time could be not only Euclidean or curved one,
but also a Cantorian space-time. In this way, accepting the new
laws of space--time distortion, one can formulate an elegant
cosmology from the point of view of the theory of groups, and be
able to unify concepts apparently not conciliatory as in the
following: the stationary and the expanding Universes, the curved
and the flat Universes, the finite and the infinite ages of the
Universe, and finite and infinite space.

The paper is organized as follows: we firstly discuss the
coordinate lines of imaginary time in Sect.2; Sect.3 presents
hyperbolic geometry and special relativity; Sect.4 is devoted to
Fantappi\`{e}'s group; in Sect.5 we consider Nuclear electro--weak
space--time and gravitational space--time, while in Sect.6
conclusions are drawn.

\section{Coordinate lines of imaginary time}

Let us start by considering the following problem. Even if, notwithstanding
relativity, we are able to define the notion of universal cosmic time, how
can we be sure that the cosmic clock has always ticked in the same way from
the beginning of time? There is no logical need, nor are there physical
theories, which answer this question. The first to explore the possibility
of the existence of more time gauges were Milne and Dirac who, however, were
not able to find a mathematics to support their ideas. Let us remember that,
in the construction of cosmological models, it is not possible to deduce the
contour conditions around the outside of the Universe.\ %@@@
We can choose many different conditions, but we need to calculate
their consequences to see if they agree with the observations.
Hartle and Hawking eliminated the problem of contour conditions
because their Universe has no frontier \cite{Hawking1},
\cite{Hawking2}. The main difficulty is to understand in what way
real time emerges continuously from imaginary time. To overcome
this difficulty, we develop the hypothesis that imaginary time is
the fundamental structure of the entire space--time manifold,
while real time does not emerge in a past age because of some
unknown physical mechanism. However, it is relevant to stress that
actually we find a quantum and relativistic imprinting as showed
in \cite{Iovane4}. It simply originates from our senses; it is
simply what we are able to perceive and measure. Along the
coordinate lines of imaginary time all the events are placed.
These events, past, present and future, simply exist in the
Universe of imaginary time. The entirety of space--time is
represented by a 4--dimensional pseudo-hypersphere\footnote{Pseudo
means in the context of Cantorian support, that is an hypersphere
with stochastic  self-similar fluctuations on the surface.}, which
exists in its entirety and is immutable. By sectioning the sphere
with planes orthogonal to the coordinate lines of imaginary time,
one sees that this model represents a stationary Universe of
cyclic imaginary time. By transforming imaginary time into real
time, we obtain the passage from a space--time pseudo-hypersphere
of imaginary time, to a pseudo-hyperboloid of real time.
Therefore, we obtain a space-time manifold of hyperbolic
structure. Among all the possible structures for space--time, the
hyperbolic is the most natural, since, as we will see in the next
paragraph, Lobacevskij--Bolyai's geometry is formally analogous to
special relativity.

\section{Hyperbolic geometry and special relativity}

Let $K$ be an inertial frame of reference and consider the hyperbole of
equation~\footnote{%
\ For sake of simplicity we are considering a 2--dimensional relativistic
space--time.}
\begin{equation}
(x^{4})^{2}-(x^{1})^{2}=R^{2}\ .
\end{equation}
The two branches of the hyperbole approach the universe lines described by
the light rays asintotically.
%, therefore, the light cone centered at the origin.
Let $K^{\prime }$ be another inertial frame of reference moving with respect
to $K,$ and let $P$ be the space--time point which is the intersection of
the upper branch of the hyperbole with the axis $x^{\prime 4}$ represented
by $x^{1}=\beta x^{4}.$ The coordinates of this point, obtained by combining
the equations of the two curves, are
\begin{equation}
\left\{
\begin{array}{l}
x^{1}=\frac{\beta R}{\sqrt{1-\beta ^{2}}}, \\
\\
x^{4}=\frac{R}{\sqrt{1-\beta ^{2}}}\ .
\end{array}
\right.
\end{equation}
By comparing Lorentz's transformations \cite{Ruffini}
\begin{equation}
\left\{
\begin{array}{l}
x^{\prime 1}=\frac{x^{1}-\beta x^{4}}{\sqrt{1-\beta ^{2}}}, \\
\\
x^{\prime 4}=\frac{x^{4}-\beta x^{1}}{\sqrt{1-\beta ^{2}}}\ ,
\end{array}
\right.
\end{equation}
one sees that the coordinates of the point represent time $R/c$ and zero
length in the primed frame of reference. For each relative velocity of $%
K^{\prime }$ with respect to $K$, and so for each inclination of the axes,
the intersection of time axis with this hyperbole will give time $R/c$.
Therefore, the hyperboles represent, in Minkowski's space--time, the locus
of the points equidistant from the origin and therefore they are the
analogous of the circumferences in the euclidian plane. By adding a spatial
dimension, one sees that the two--sheeted hyperboloids are the analogous of
the spheres and similarly, by adding other spatial dimensions, we can
construct the hyperspheres of relativistic geometry. These spheres in the
pseudo--euclidian spaces have been amply studied by mathematicians and it
has been demonstrated that their intrinsic geometry is Lobacevski's
hyperbolic geometry. Precisely, they are surfaces of negative constant
curvature $K=-1/R^{2}$.

We see from this that hyperbolic geometry, in respect to euclidian geometry
and elliptic geometry, has a particularly important role in space--time, and
from the point of view of group theories, is very similar to the theory of
relativity where, in space--time, relativity and hyperbolic geometry share
Poincar\`{e}'s symmetry group.

\section{Fantappi\`{e}'s group}

Fantappi\`{e} noted that general relativity follows an extraneous approach
to the tradition of mathematical physics in that it does not follow the
group structure of physics. This is different from classical mechanics and
special relativity.

Let us remember that Galileo's group is the main group of classical physics
and is formed by the composition of the following transformations: \newline
a) Spatial Rotations - characterized by three parameters
\begin{equation}
x_{\mu }^{\prime }=a_{\mu \nu }x_{\nu }\quad ,\quad t^{\prime }=t,
\end{equation}
where $[a_{\mu \nu }]$ is an orthogonal matrix whose determinant
is +1.
\newline
b) Inertial Movements - characterized by the three components of
velocity,
\begin{equation}
x_{\mu }^{\prime }=x_{\mu }+v_{\mu }t\quad ,\quad t^{\prime }=t.
\end{equation}
c) Spatial translations - characterized by three parameters,
\begin{equation}
x_{\mu }^{\prime }=x_{\mu }+a_{\mu }\quad ,\quad t^{\prime }=t.
\end{equation}
d) Temporal translations - characterized by only one parameter,
\begin{equation}
x_{\mu }^{\prime }=x_{\mu }\quad ,\quad t^{\prime }=t+t_{0}.
\end{equation}
Therefore Galileo`s group has order 10 and expresses Galileo's well-known
relativity principle. Moving on to relativistic physics, spatial rotations
and inertial movements become fused in a unique operation, the rotations of
a euclidian space $M_{4}$, characterized by 6 parameters,
\begin{equation}
x_{i}^{\prime }=a_{ik}x_{k},
\end{equation}
where $\left| a_{ik}\right| =1,$ $x_{1}=x$, $x_{2}=y$, $x_{3}=z$, $x_{4}=ict$%
.

These transformations, called Lorentz's special transformations, form
Lorentz's proper group and joining the reflections, form Lorentz's extended
group. Then we need to add the translations of $M_{4}$
\begin{equation}
x_{i}^{\prime }=x_{i}+a_{i},
\end{equation}
characterized by 4 parameters, which comprise spatial and temporal
translations. By composing the transformations of these two groups, we
obtain Lorentz's general transformations which form Poincar\'{e}'s group of
10 parameters
\begin{equation}
x_{i}^{\prime }=a_{ik}x_{k}+a_{i}\ .
\end{equation}
Poincar\`{e}'s group mathematically translates Einstein's relativity
principle. When $c\rightarrow \infty $ so that $\frac{v}{c}\ll 1$,
Minkowski's space--time reduces to that of Newton's and Poincar\`{e}'s group
reduces to Galileo's group.

Fantappi\`{e} went on in this direction and tried to understand if
Poincar\`{e}'s group could be the limit of a more general group,
in the same manner as Galileo's group is the limit of
Poincar\`{e}'s group. In \cite {Fantappie} He wrote a new group of
transformations, which had as limit Poincar\`{e}'s group and He
was able also to demonstrate that his group was not able to be the
limit of any continuous group of 10 parameters. That is, by
limiting to groups of 10 parameters and to 4--dimensional spaces,
what happened with Galileo's and Poincar\`{e}'s groups cannot be
repeated. For this reason this group is called the final group.

Fantappi\`{e}, moving from a space--time with hyperbolic structure, showed
that, through a flat projective representation, one could obtain a
space--time which generalizes Minkowski's space--time.

Let us remember that to have a flat representation of hyperbolic
geometry, we fix a circle in the plane, with center $O$ and radius
$r$, called the absolute of Cayley-Klein. Relative to this we get
the following definitions:

\begin{equation*}
\left\{
\begin{array}{lll}
\text{point} & \Rightarrow &
\begin{array}{l}
\text{point inside the circle;}
\end{array}
\\
\text{straight line} & \Rightarrow &
\begin{array}{l}
\text{chord of the circle (without extremes);}
\end{array}
\\
\text{plane} & \Rightarrow &
\begin{array}{l}
\text{region of points inside the circle;}
\end{array}
\\
\text{movements} & \Rightarrow &
\begin{array}{l}
\text{projections on the plane that transform} \\
\text{the region of the internal points in itself;}
\end{array}
\\
\text{congruent figures} & \Rightarrow &
\begin{array}{l}
\text{figures which can be transformed from} \\
\text{one to the other through a projection.}
\end{array}
\end{array}
\right.
\end{equation*}
Let us introduce a system of orthogonal coordinates with origin in the
center of the circle. It is not possible to represent the distance of two
points $A(x,y)$ and $B(x^{\prime },y^{\prime })$ in the form $\sqrt{%
(x-x^{\prime })^{2}+(y-y^{\prime })^{2}}$ \ because it is not invariant for
the projective transformations. An expression of the coordinates of $A$ and $%
B$, which remains invariable for all the projective transformations which
leave the limit circle fixed, is the anharmonic ratio of the four points $%
A,B,M,N$
\begin{equation}
(ABMN)=(AM/BM):(BN/AN),
\end{equation}
where $M$ and $N$ are the extremes of chord $AB.$ \newline
We assume as distance
\begin{equation}
\mathrm{dist}(AB)=k\log (ABMN).
\end{equation}
In this way one sees that every point of the hyperbolic plane, no matter how
they are moved, always remain at infinite distance from the points of the
\emph{absolute}. So the hyperbolic plane is finite and limited if we make
the measures in a euclidian sense. On the contrary, if the measures are not
euclidian, the hyperbolic plane is infinite and unlimited.

Fantappi\`{e} choses, as the absolute quadric, the hypersphere
\begin{equation}
x_{E}^{2}+y_{E}^{2}+z_{E}^{2}-c^{2}t_{E}^{2}+R^{2}=0,
\end{equation}
which, in the 2--dimensional case becomes the circumference
\begin{equation}
x_{E}^{2}-c^{2}t_{E}^{2}+R^{2}=0\ .  \tag{13'}
\end{equation}
He showed that Minkowski's space--time can be considered as a limit case of
the projected space--time when $R\rightarrow \infty $. Therefore
Poincar\`{e}'s group proves the limit of the group of motions of new
space--time in itself. To determine the transformations of the new group, we
observe that the space--time motions are represented by the projections that
transform the absolute circumference in itself. Such absolute, with the
introduction of imaginary time, can be written as
\begin{equation}
x_{1}^{2}+x_{2}^{2}+x_{3}^{2}+x_{4}^{2}+R^{2}=0
\end{equation}
and, by considering the homogeneous coordinates $\overline{x}%
_{A}(A=1,2,..,5) $ so defined\footnote{%
\ Let us remember that if $(x_{0},y_{0},z_{0})$ are the cartesian orthogonal
coordinates of a point $P$ in ordinary space, one defines the four
homogeneous coordinates by the relations
\begin{equation*}
x_{0}=\frac{x_{1}}{x_{4}}\qquad y_{0}=\frac{x_{2}}{x_{4}}\qquad z_{0}=\frac{%
x_{3}}{x_{4}}\ .
\end{equation*}
\begin{equation*}
\mbox{If}\qquad Ax^{2}+By^{2}+Cz^{2}+Dxy+Exz+Fyz+Gx+Hy+Iz+L=0
\end{equation*}
is the equation of a quadric in cartesian orthogonal coordinates, then in
homogeneous coordinates we have
\begin{equation*}
Ax_{1}^{2}+Bx_{2}^{2}+Cx_{3}^{2}+Dx_{1}x_{2}+Ex_{1}x_{3}+Fx_{2}x_{3}+Gx_{1}x_{4}+Hx_{2}x_{4}+Ix_{3}x_{4}+Lx_{4}^{2}=0.
\end{equation*}
}
\begin{equation}
x_{k}=R\overline{x}_{k}/\overline{x}_{5}\ ,
\end{equation}
it becomes
\begin{equation}
\overline{x}_{1}^{2}+\overline{x}_{2}^{2}+\overline{x}_{3}^{2}+\overline{x}%
_{4}^{2}+\overline{x}_{5}^{2}=0\ .
\end{equation}
It follows that the motions we are searching for are those which leave the
prior quadratic form invariable and are the orthogonal substitutions on the
five variables $\overline{x}_{A}.$ These transformations form the group of
5--dimensional rotations and are three types (moving to the non--homogeneous
coordinates):

a) \textit{\ Time translations}: considering two observers standing still in
the same place, but separated by a great distance in time, that is, the same
observer in two different moments
\begin{equation}
\left\{
\begin{array}{l}
x^{\prime }=\frac{x\sqrt{1-\eta ^{2}}}{1-\eta t/(R/c)}, \\
\\
t^{\prime }=\frac{t-T}{1-\eta t/(R/c)},
\end{array}
\right.
\end{equation}
with $\eta =\frac{T}{R/c}$ where $T$ is the parameter of time translation.
It follows that for $t=\pm R/c$, one has $x^{\prime }=0$.

These transformations for $R\rightarrow \infty $ are reduced to the classic
time translations
\begin{equation}
x^{\prime }=x\qquad ,\qquad t^{\prime }=t-T
\end{equation}
\qquad b) \textit{Spatial Translations}: considering two observers at the
same time and standing still compared to each other, but separated by great
distance in space (for example along the $x$ axis)
\begin{equation}
\left\{
\begin{array}{c}
x^{\prime }=\frac{x-S}{1+\alpha x/R}, \\
\\
y^{\prime }=\frac{y\sqrt{1+\alpha ^{2}}}{1+\alpha x/R}, \\
\\
z^{\prime }=\frac{z\sqrt{1+\alpha ^{2}}}{1+\alpha x/R}, \\
\\
t^{\prime }=\frac{t\sqrt{1+\alpha ^{2}}}{1+\alpha x/R},
\end{array}
\right.   \label{spatialtransations}
\end{equation}
with $\alpha =\frac{S}{R}$ and where $S$ is the parameter of translation
along the $x$ axis. \newline
At the relativistic limit, that is for $R\rightarrow \infty $, equations (%
\ref{spatialtransations}) reduce to
\begin{equation}
x^{\prime }=x-S,\qquad y^{\prime }=y,\qquad z^{\prime }=z,\qquad t^{\prime
}=t\ .
\end{equation}
c) \textit{Pullings}: considering two observers that initially coincide, and
one moving rectilinearly and uniformly to the other, with velocity parallel
to the $x$ axis
\begin{equation}
\left\{
\begin{array}{l}
x^{\prime }=\frac{x-Vt}{\sqrt{1-\beta ^{2}}}, \\
\\
y^{\prime }=y, \\
\\
z^{\prime }=z, \\
\\
t^{\prime }=\frac{t-Vx/c^{2}}{\sqrt{1-\beta ^{2}}}\ .
\end{array}
\right.
\end{equation}
Let us synthesize the geometric structure of Fantappi\`{e}'s group in the
following scheme:

\begin{equation*}
\begin{array}{lll}
\text{GALILEO'S GROUP} & \Rightarrow & \text{Rotations\ S}_{3} \\
& \Rightarrow & \text{Pullings} \\
& \Rightarrow & \text{Translations\ S}_{3} \\
& \Rightarrow & \text{Time\ Translations} \\
&  &  \\
\text{POINCARE'S GROUP} & \Rightarrow & \text{Rotations\ S}_{4} \\
& \Rightarrow & \text{Translations\ S}_{4} \\
&  &  \\
\text{FANTAPPIE'S GROUP} & \Rightarrow & \text{Rotations\ S}_{5} \\
&  &
\end{array}
\end{equation*}

\section{Nuclear electro--weak space--time and gravitational space--time}

Some gravitational phenomena show us that in a curved space and in
some conditions, we cannot localize an object in its effective
position (this is the case of the gravitational lenses). Instead
the source is seen by the observer in the direction of the tangent
to the light rays in the point where we are \cite{Falco}, and as a
curved space appearing as if it were flat. It could seem likely
that something like this can also happen for time. It could seem
that the entire space--time manifold, in which every observer can
watch the phenomena, is only a flat representation of the
space--time tangent to curved manifold. Applying the flat
representation of hyperbolic geometry, the group of motions of the
space--time manifold in itself is represented by Fantappi\`{e}'s
group. At this point, the idea is that there is a difference
between the projected space--time $(x_{E},t_{E})$ that every
observer can see and is associated with atomic processes and with
light frequencies, and the non--projected space--time
$(x_{G},t_{G})$ associated with gravitational phenomena. The
projected space--time coordinates are regulated by Fantappi\`{e}'s
transformations and are on a Cantorian support. Let us determine,
therefore, the relations that link projected and non--projected
coordinates together. Let us consider the hyperbolic model with
real time and let us make a flat representation of it by choosing
as \emph{absolute} the hypersphere
\begin{equation}
\ c^{2}t_{E}^{2}-x_{E}^{2}-y_{E}^{2}-z_{E}^{2}-R^{2}=0,
\end{equation}
which, to simplify, in the 2--dimensional case becomes the circumference
\begin{equation}
c^{2}t_{E}^{2}-x_{E}^{2}=R^{2}\ .  \tag{22'}
\end{equation}
Let us calculate the time distance of two points, $A(0)$ and $B(t_{E})$,
placed on the $t_{E}$ axis. The time axis meets the absolute at two points $%
C(R/c)$ and $D(-R/c)$ and so
\begin{equation}
(ABCD)=\frac{AC}{CB}\frac{DB}{AD}=\frac{C-A}{B-C}\frac{B-D}{D-A}=\frac{R/c}{%
t_{E}-R/c}\frac{t_{E}+R/c}{-R/c}=\frac{R+ct_{E}}{R-ct_{E}}\ .
\end{equation}
The measure of the time interval between the two events on the
$t_{E}$-axis is
\begin{equation}
t_{G}=\frac{R}{c}\log \frac{R+ct_{E}}{R-ct_{E}}\ ,
\end{equation}
by posing $k=\frac{R}{c}$.

So we conclude that time $t_{E}$, linked to the non--gravitational
interactions, is slower than gravitational time. Taking into account the
observed data (current radius of the Universe, speed of light, etc.) one can
easily verify that we have to consider intervals of many thousands of years
in such a way that the two temporal scales differ by only a second. Going
back instead to a past cosmologic epoch, the differences increase and when
electro--magnetic time nears $-t_{EU}=R/c$, which for us coincides with the
beginning of time, gravitational time extends into the infinite past. That
is, in gravitational time, the Universe is infinitely old. Obviously every
observer sees only a part of the manifold and that is what we call the past.
So, as the background cosmic radiation demonstrates, there was an initial
instant in respect to the electro--magnetic waves, while the Universe is
eternal in the gravitational scale. By considering two points placed on the $%
x_{E}$-axis, $A(0)$ and $B(x_{E}),$ we will have, instead, $P(-iR)$ and $%
Q(iR)$, where their spatial distance will be
\begin{equation}
x_{G}=\frac{R}{2i}\log \frac{iR-x_{E}}{x_{E}+iR}=\frac{R}{2i}\log \frac{%
-R-ix_{E}}{ix_{E}-R}=\frac{R}{2i}\log \frac{R+ix_{E}}{R-ix_{E}}=Rarctg\frac{%
x_{E}}{R},
\end{equation}
by posing $k=\frac{R}{2i}$.

In other words, non--projected space is smaller than projected one.

From the prior relations the two inverse formulas follow:
\begin{equation}
\left\{
\begin{array}{l}
t_{E}=\frac{R}{2c}tgh\frac{ct_{G}}{R}, \\
\\
x_{E}=Rtg\frac{x_{G}}{R}\ .
\end{array}
\right.
\end{equation}
In conclusion, therefore, in cosmologic times, the astronomic clocks slowly
lose their synchronization with the atomic clocks.

In gravitational time Universe space is finite and time is infinite; in a
projected Universe space is infinite and time is finite.

This physical interpretation of Fantappi\`{e}'s transformations implies that
the electromagnetic age of the Universe is constant. The temporal
translations demonstrate
\begin{equation}
t_{E}=\frac{t_{E1}+t_{E2}}{1+t_{E1}t_{E2}/t_{EU}^{2}}\ .
\end{equation}
This relation is equal in form to relativistic law of the composition of
velocities. Therefore, as such, the speed of light is the same for each
observer in whichever motion, and as being finite, cannot be exceeded. In
the same way the electromagnetic age of the Universe is the same for each
observer, whichever its space--time position. Every observer will see the
same Universe globally, not only from every point of space, but also in any
era. This is not different from the perfect cosmological principle
postulated by the authors of the stationary model, who however, had to
hypothesize the creation of new matter from nothing in order to verify it.
In this treatment the perfect cosmological principle can be obtained as a
consequence of Fantappi\`{e}'s group.

Differentiating Fantappi\`{e}'s temporal translations, we obtain
\begin{equation}
dx_{E}^{\prime }=\frac{\sqrt{1-\eta ^{2}}}{1+\eta t_{E}/t_{EU}}dx_{E}-\frac{%
x_{E}\sqrt{1-\eta ^{2}}}{(1+\eta t_{E}/t_{EU})^{2}}\frac{\eta }{t_{EU}}%
dt_{E},
\end{equation}
\begin{equation}
dt_{E}^{\prime }=\frac{1+\eta t_{E}/t_{EU}-(t_{E}+T_{E0})\eta /t_{EU}}{%
(1+\eta t_{E}/t_{EU})^{2}}dt_{E}=\frac{1-\eta ^{2}}{(1+\eta t_{E}/t_{EU})^{2}%
}dt_{E},
\end{equation}
and so
\begin{equation}
V_{E}^{\prime }\sqrt{1-\eta ^{2}}=V_{E}(1+\eta t_{E}/t_{EU})-x_{E}\eta
/t_{EU},
\end{equation}
which furnishes the link between the velocities of a point measured in two
instants of electromagnetic time separated by the interval $T_{E0}.$

In particular, if the interval of time is $T_{E0}=t_{EU}=R/c$, since $\eta
=1 $, we obtain,
\begin{equation}
V_{E}=\frac{x_{E}}{t_{E}+t_{EU}}=H(t_{E})x_{E}\ .
\end{equation}
At the current time, that is $t_{E}=0,$ we have
\begin{equation}
H=1/t_{EU}=c/R\ .
\end{equation}
Therefore our physical interpretation of Fantappi\`{e}'s transformations
says that every observer will see an expanding Universe with escape velocity
proportional to the distance, and this agrees with Hubble's law.

\section{Physical Remarks and Conclusions}

We know that in order, for a body in rotation on itself, to be
dynamically stable there must be a condition of equilibrium
between gravitational force, which depends on its mass, and
centrifugal force, which depends on its velocity of rotation. If
the body rotates faster than a certain maximum velocity, it will
disintegrate because of its own centrifugal force. Observations of
stellar objects with highly elevated velocities of rotations,
actually show periods of rotation of the order of a few
milliseconds. To withstand the centrifugal force, they should have
a density of the order of $10^{14}gr/cm^{3}.$ Such high density is
equal only to that of an atomic nucleus, so for this reason
astrophysicists think these objects must be neutron stars, but it
could be another answer: they could stay on a Cantorian space-time
and our view is just a projection.

There are galaxies which rotate faster than the theoretical
maximum velocity and beyond this the velocity of the stars along
the arms do not seem to decrease in a keplerian way. To justify
the equilibrium one hypothesizes the existence of dark matter
which increases the mass. However, the study of space--time
through Fantappi\`{e}'s group acknowledges a new law of time
dilatation that we associated with the two different time scales.
That is, it seen through electromagnetic time, processes such as
the rotation of celestial bodies were strongly accelerated in the
past, while the behavior of light and atomic processes remained
invariable. Therefore, even in this context, the application of
Fantappi\`{e}'s transformations agrees with the observations. It
is obvious that, for small space--time scales, the results are not
different for $t_{G}$ in respect to $t_{E.}$

The results of our paper are not conclusive and just put into
evidence that the cosmological problems and Fantappi\`{e}'s
physical--mathematical observations, based on the theory of
groups, can be framed in the same interpretive scheme. It is
particularly interesting to observe that projective geometry on
Cantorian space-time behaves as natural geometry in cosmology. In
addition, we have shown how Hartle and Hawking's ideas of
imaginary time, extended to the entire Cantorian Universe, allow
us to see the model of Big Bang and the stationary model, as two
different projections of the same reality. In particular, the two
models appear to be connected to two different time scales; the
latter linked to space--time geometry, and therefore to
gravitational interaction; the prior is linked to the other
fundamental interactions which measure time for each observer.

\subsubsection*{Acknowledgements}

The authors wish to thank E.Benedetto for comments and
discussions.

\end{document}